\documentclass[11pt,a4paper]{article}

\usepackage{amsfonts,amsmath,amssymb}
\usepackage{mathrsfs}
\usepackage{amscd}
\usepackage{color}
\usepackage{graphicx}

\font\tengot=eufm10  \font\sevengot=eufm7  \font\fivegot=eufm5

\newfam\gotfam
\textfont\gotfam=\tengot
\scriptfont\gotfam=\sevengot
\scriptscriptfont\gotfam=\fivegot

\begin{document}
	
\title {Non-spherical sources of Schwarzschild space--time}

\author{J.L. Hern\'andez--Pastora\thanks{Departamento de Matem\'atica Aplicada. Facultad de Ciencias. Universidad de Salamanca,  and  IUFFyM. https://ror.org/02f40zc51.  e-mail address: jlhp@usal.es. ORCID:orcig.org/0000-0002-3958-6083} \ and  L. Herrera \thanks{Instituto Universitario de F\'\i sica Fundamental y Matem\'aticas (IUFFyM) e-mail address: lherrera@usal.es} \\}

\date

\maketitle

\vspace{-10mm}

\begin{abstract}
While it is known that any spherical fluid distribution  may only source the spherically symmetric Schwarzschild space--time, the inverse is not true. Thus, in this manuscript,  we find exact axially symmetric and static fluid (interior) solutions to Einstein equations, which match smoothly on the boundary surface to the Schwarzschild  (exterior) space--time, even though the fluid distribution is not endowed with spherical symmetry. The solutions are obtained  by using the general approach outlined in  \cite{weylsol}, and satisfy the usual requirements imposed to any physically admissible interior solution. A discussion about the physical and geometric properties of the source is presented. The relativistic multipole moments ($RMM$) are explicitly  calculated in terms of the physical variables, allowing to prove that spherical sources can only match to the Schwarzschild space--time. The complexity of the source is evaluated through the complexity factors. It is shown that there is only one independent complexity factor, as in the spherically symmetric case.
 \end{abstract}


\large


\section{Introduction}
It is a common conjecture in the modeling of compact stars, within the context of general relativity, to assume that the fluid distribution  describing the source is endowed with the same symmetries as the exterior solution it produces. However we know by some  counterexamples that such an assumption is, in general,  incorrect. The  best known counterexample to the above mentioned conjecture is provided by the Szekeres  space--time \cite{1,2}. This solution represents  dust models which have no Killing vectors (not even a time--like one) \cite{3,5}, and still may be matched smoothly to the Schwarzschild line element \cite{3}, a result which allowed  Bonnor to conclude that such space--time does not produce gravitational radiation.

It is the purpose of this work, to find a family of solutions representing non--spherical, static,  fluid distributions which match smoothly to the Schwarzschild line element. 
The solutions are obtained by using the general approach outlined in \cite{weylsol}, and satisfy the usual requirements imposed to any physically admissible interior solution. 

In \cite{epjc_vol}, both  anisotropic and isotropic spherical interior sources, as well as non-spherical ones, smoothly matched  to  any exterior Weyl solution were obtained. The metric functions of the interior solution for a global model could be integrated in terms of some combination of the energy--momentum tensor components, thereby providing some constraints  
on the source,  derived from the
exterior gravitational field of the global metric. Thus, these constraints
can be expressed in terms of the gravitational field
which is matching the interior solution. The fact that  the method
proposed in \cite{weylsol} allows us to construct a well matched interior
metric for any exterior solution of the Weyl family, makes it possible to  find out,  for any vacuum solution, how the energy--momentum tensor of axially symmetric static sources is
affected by different physical characteristics of the gravitational field outside the source.

The ideas briefly exposed above   lay the foundations of the method to obtain the solutions we are looking for. Such a method shall be analyzed in detail in next sections. 

The physical and geometric properties of the obtained source are analyzed in detail. The trace--free part of the electric Riemann tensor is calculated. It is shown that it has only one independent component, implying that the source is characterized by a single complexity factor, as in the spherically symmetric case.

Finally, let us recall that  the relativistic multipole moments ($RMM$) \cite{geroch} are scalars which asymptotically characterize any vacuum stationary solutions to the Einstein equations. When considering global solutions (in and out of the source) it becomes necessary to relate the  $RMM$ to  the matter content of the source.  Such a link has been established in 
 \cite{whom}, through explicit expressions of the  $RMM$ in terms of integrals over the space--time filled by the source.  In other words, for any interior solution smoothly matched to any vacuum metric, it is possible to calculate the corresponding $RMM$, which of course correspond to the exterior vacuum solution. Accordingly, any interior solution  smoothly matched  to the Schwarzschild space--time,  is characterized by a single multipole moment (the monopole), all other multipole moments vanishing.  This result allows us to prove that although the source of the Schwarzschild line element may be a non--spherical source, any (static) spherical source may only produce a Schwarzschild space--time.

\section{The interior metric and sources}

We shall first  briefly summarize  the general method developed in \cite{weylsol}  to find matchable solutions to the Weyl space--time, and describe the general conventions and notation.
 
\subsection{The global model}

In  \cite{weylsol} a global metric is obtained for any static solution belonging to the Weyl family of the axisymmetric vacuum Einstein equations. The exterior line element written in Erez-Rosen coordinates is given by 
\begin{eqnarray}
ds^2_E&=&-e^{2\psi(r,y)}dt^2+e^{-2\psi+2\left[\Gamma(r,y)-\Gamma^s
	\right]}dr^2+
e^{-2\left[\psi-\psi^s\right]+
	2\left[\Gamma(r,y)-\Gamma^s\right]}r^2 d\theta^2\nonumber \\
&+& e^{-2\left[\psi-\psi^s\right]} r^2\sin^2\theta d\phi^2,
\label{exterior}
\end{eqnarray}
where  $\psi^s$ and $\Gamma^s$ are the metric functions corresponding to the Schwarzschild solution, namely,
\begin{equation}
\psi^s=\frac 12 \ln \left(\frac{r-2M}{r}\right) \, \quad \Gamma^s=-\frac 12 \ln \left[\frac{(r-M)^2-y^2M^2}{r(r-2M)}\right],
\label{schwfun}
\end{equation}
 the parameter $M$ being easily identified as the Schwarzschild mass. 
 
For the interior axially symmetric line element we shall assume

\begin{equation}
ds^2_I=-e^{2 \hat a} Z(r)^2 dt^2+\frac{e^{2\hat g-2\hat a}}{A(r)} dr^2+e^{2\hat g-2\hat a}r^2 d\theta^2+e^{-2 \hat a}r^2 \sin^2\theta d\phi^2,
\label{interior}
\end{equation}
with $A(r)\equiv 1-pr^2$  and  $Z\equiv\displaystyle{ \frac 32 \sqrt{A(r_{\Sigma})}-\frac 12 \sqrt{ A(r)}}$, and where $p$ is an arbitrary constant and the boundary surface of the source is defined by $r=r_{\Sigma}=const.$ Both functions correspond to the well known incompressible (homogeneous energy density) perfect fluid sphere in the limiting case $\hat g=\hat a=0$, and the matching conditions with the Schwarzschild solution imply $p=\displaystyle{\frac{2M}{r_{\Sigma}^3}}$, 
where the other interior metric functions have the following form:
\begin{eqnarray}
\hat a(r,\theta)&=&\hat \psi_{\Sigma} s^2(3-2s)   +r_{\Sigma}\hat \psi^{\prime}_{\Sigma}s^2(s-1)+(r-r_{\Sigma})^2F(r,\theta),\nonumber \\
\hat g(r,\theta)&=&\hat \Gamma_{\Sigma} s^3(4-3s)   +r_{\Sigma}\hat \Gamma^{\prime}_{\Sigma}s^3(s-1)+(r-r_{\Sigma})^2G(r,\theta).
\label{aygsimple}
\end{eqnarray}
with $s\equiv r/r_{\Sigma} \in \left[0,1\right]$, $\hat \psi \equiv \psi-\psi^s$ and  $F(r,\theta)$, $G(r, \theta)$ arbitrary functions satisfying  the following constraints: $F(0,\theta)=F^{\prime}(0,\theta)=0$,  $G(0,\theta)=G^{\prime}(0,\theta)=G^{\prime \prime}(0,\theta)=0$ derived from the matching conditions and regular behavior at the origin (prime denotes derivative with respect to $r$).

Thus, for our line element  (\ref{interior}) we have the following non vanishing components of the energy--momentum tensor (see \cite{weylsol} for details)
\begin{eqnarray}
-T^0_0&=&\gamma \left(8 \pi \mu+\hat p_{zz}-E\right),\nonumber\\
T^1_1&=& \gamma \left(8 \pi P-\hat p_{xx}\right),\nonumber \\
T^2_2&=& \gamma \left(8 \pi P+\hat p_{xx}\right),\nonumber \\
T^3_3&=&\gamma \left(8 \pi P-\hat p_{zz}\right),
\label{eegeneral}
\end{eqnarray}
	\begin{equation}
	T_1^2=g^{\theta\theta}T_{12}=-\frac{\gamma}{r^2}\left[2 {\hat a}_{,\theta}  \hat a^{\prime}-\hat g^{\prime}\frac{\cos\theta}{\sin\theta}-\frac{\hat g_{,\theta}}{r}+\frac{(1-A)}{r \sqrt A (3 \sqrt{A_{\Sigma}}- \sqrt{A})}(2 {\hat a}_{,\theta} -{\hat g}_{,\theta}  )\right],
	\label{pxy}
	\end{equation}
with $\displaystyle{\gamma\equiv \frac{e^{2\hat a-2\hat g}}{8 \pi}}$, subscript denotes derivative with respect to the angular variable, and
\begin{eqnarray}
&&E=-2 \Delta \hat a+(1-A)\left[2 \frac{\hat a^{\prime}}{r}\frac{9 \sqrt{A_{\Sigma}}-4 \sqrt{A}}{3 \sqrt{A_{\Sigma}}- \sqrt{A}}+2 \hat a^{\prime \prime}\right],\nonumber\\
&&\Delta \hat a= \hat a^{\prime \prime}+2\frac{\hat a^{\prime}}{r}+\frac{{\hat a}_{,\theta \theta} }{r^2}+\frac{{\hat a}_{,\theta} }{r^2}\frac{\cos \theta}{\sin \theta},\nonumber \\
&&\hat p_{xx}=-\frac{{\hat a}_{,\theta} ^2}{r^2}-\frac{\hat g^{\prime}}{r}+\hat a^{\prime 2}+\frac{{\hat g}_{,\theta} }{r^2}\frac{\cos \theta}{\sin \theta}+\nonumber\\
&+&(1-A)\left[2 \frac{\hat a^{\prime}}{r}\frac{\sqrt{A}}{3 \sqrt{A_{\Sigma}}- \sqrt{A}}- \hat a^{\prime 2} +\frac{\hat g^{\prime}}{r}\frac{3 \sqrt{A_{\Sigma}}-2 \sqrt{A}}{3 \sqrt{A_{\Sigma}}-\sqrt A}\right], \nonumber \\
&&\hat p_{zz}=-\frac{{\hat a}^2_{,\theta} }{r^2}-\frac{\hat g^{\prime}}{r}-\hat a^{\prime 2}-\frac{{\hat g}_{,\theta \theta} }{r^2}-\hat g^{\prime \prime} A+\nonumber\\
&+&(1-A)\left[-2 \frac{\hat a^{\prime}}{r}\frac{\sqrt{A}}{3 \sqrt{A_{\Sigma}}- \sqrt{A}}+ \hat a^{\prime 2} +2\frac{\hat g^{\prime}}{r}\right],
\label{eegeneraldet}
\end{eqnarray}
where $\hat p_{zz}, \hat p_{xx}, T_1^2, E$ describe deviations from the spherical symmetry.
Indeed, if $\hat a=\hat g=0$,  then $E=\hat p_{xx}=\hat p_{zz}=0$ and we recover the  spherical case of an incompressible perfect fluid sphere with \begin{eqnarray}
-T^0_0\equiv \mu&=&\frac{3}{4 \pi \tau r_{\Sigma}^2 },\nonumber\\
T^1_1=T^2_2=T^3_3\equiv P&=&\mu\left(\frac{\sqrt A-\sqrt{A_{\Sigma}}}{3 \sqrt{A_{\Sigma}}-\sqrt{A}}\right),
\label{presionP}
\end{eqnarray}
where $\tau\equiv r_\Sigma/M$ denotes the inverse of the compression factor.

\section{Non--spherical sources matched to the Schwarzschild space--time}

We shall now tackle the problem motivating this work. If we consider that the exterior space--time is  spherically symmetric (Schwarzschild), then the interior metric functions (\ref{aygsimple}) would be
\begin{equation}
\hat a(r,\theta)=(r-r_{\Sigma})^2F(r,\theta),\qquad 
\hat g(r,\theta)=(r-r_{\Sigma})^2G(r,\theta),
\label{aygsimpleschw}
\end{equation}
which are  arbitrary,  with the only constraints mentioned above for $F$ and $G$ with respect to their behaviour at the origin of coordinates as well as conditions preserving the good behaviour at the symmetry axis $G(r,\theta=0)=0$. In addition both functions should lead to expressions of  the energy-momentum tensor of the source whose components must fulfill the usual  energy conditions

{\it i)} positive radial pressure $g_{11}T_1^1\equiv P_r >0$, 

{\it ii)} the strong energy condition (S.E.C.) $(-T_0^0)-T_i^i>0$ and 

{\it ii)} the positive energy density (P.E.D.) $ -T_0^0>0$.

\vskip 2mm
\subsection{Spherical sources}
In order to illustrate our approach, it could be convenient to  describe briefly the case of spherical sources. In  such a case,  the arbitrary functions $F$ and $G$ only depend on the radial coordinate, and we are able to recover the results obtained in \cite{luisJA}, \cite{Lake} where a procedure for obtaining all the interior solutions for the spherical case has been established. 

Thus, within the spherical case the metric function $G(r)$ must be chosen to be null for two reasons: on the one hand because we want to avoid infinite pressure at the symmetry axis
($\theta = 0$), or in other words we must consider a vanishing
component $T_1^2$ of the energy--momentum tensor, and hence from (\ref{pxy}) the spherical case implies $\hat g^{\prime}= 0$, which together with the matching conditions is equivalent to $\hat g = 0$ (see \cite{epjc_vol}, \cite{whom}, \cite{weylsol} for details). 

On the other hand we want to preserve the canonical spherical form of the angular part  in the line element (in order to recover the results in \cite{luisJA}) and then we must consider $\hat g=0$ in our interior metric (\ref{interior}) as well as a change of coordinates in the radial variable, which preserve the angular part of the metric in
the form $\hat r^2 d\theta^2 + \hat r^2 \sin^2 \theta d\varphi^2$ , ${\displaystyle \hat r=re^{-\hat a(r)}}$.

In addition, for this case we have that $p_{xx} =- p_{zz}$ and then $T^2_2 = T^3_3$, which means that only two independent main stresses  exist in this case. These are  usually denoted in the literature as $p_r$ (radial pressure) and $p_{\bot}$ (tangential pressure) whenever the above spherical gauge is used for the coordinates. 

The relationship for the metric functions and the  anisotropy $\Pi(r)$ is the following:
\begin{eqnarray}
e^{\nu[\hat r=\hat r(r)]}&=&Z^2 e^{2\hat a}, \nonumber   \\
e^{-\lambda(\hat r=\hat r(r))}&=&A (1-r\hat a^{\prime})^2, \nonumber   \\
\Pi(r)&\equiv &8\pi (T_1^1-T_2^2)  \equiv 8\pi(p_r-p_{\bot})=-Ae^{2\hat a} \hat a^{\prime 2}+(1-A)2\hat a^{\prime}\frac{\sqrt A e^{2\hat a}}{3\sqrt{A_{\Sigma}}-\sqrt A}, \nonumber
\label{rela}
\end{eqnarray}
where $\nu$ and $\lambda$ denote the two spherical metric functions appearing in \cite{luisJA} $g_{00}=-e^{\nu}$ and $g_{11}=e^{\lambda}$.

\vskip 2mm
\subsection{Non-spherical sources}
Let us now turn to  the general non-spherical case,  for which  the expressions are more complicated. Thus, in order to specify our model we need  to introduce some simplifying assumptions. 

We shall assume  $\hat a=0$. Such a choice is justified by the fact that, as can be seen from (\ref{eegeneraldet}), the function $E$ only depends on the metric function $\hat a$ and furthermore it implies $F(r, \theta)=0$. Hence, from (\ref{eegeneral}) the above energy conditions ({\it i)}  to {\it iii)} are all satisfied if:
\begin{equation}
p_{\pm} \equiv p_{zz}\pm p_{xx} \geq 0,\quad p_{zz} \geq 0, \quad p_{xx} \leq 0.
\label{condifunc}
\end{equation}
Besides,  because both $p_{\pm}$ and $p_{zz}$ can also be negative or $p_{xx}$ positive,  those conditions remain fulfilled if and only if
\begin{equation}
p_{xx} \in \left( \right. - \infty , \frac{6}{\tau r_{\Sigma}^2} c_s(\tau)\left. \right], \quad p_{zz} \in \left[ \right. - \frac{6}{\tau r_{\Sigma}^2} c_i(\tau), \infty \left. \right), \quad p_{\pm} \in \left[ \right. - \frac{12}{\tau r_{\Sigma}^2} c_i(\tau), \infty \left. \right),
\label{domains}
\end{equation}   
where 
\begin{equation}
c_s(\tau)\equiv \frac{\sqrt{\tau-2s^2}-\sqrt{\tau-2}}{3 \sqrt{\tau-2}-\sqrt{\tau-2s^2}} , \quad c_i(\tau)\equiv \frac{2 \sqrt{\tau-2}-\sqrt{\tau-2s^2}}{3 \sqrt{\tau-2}-\sqrt{\tau-2s^2}}.
\end{equation}
Equivalently, more restricted intervals required to verify all the energy conditions can be simply written as follows
\begin{equation}
\rvert p_{\pm}\rvert\leq 8 \pi(\mu-P) ,\quad \rvert p_{zz}\rvert \leq 4 \pi(\mu-P), \quad \rvert p_{xx}\rvert \leq 8 \pi P.
\label{condialter}
\end{equation}

In addition it is worth noticing that from the expression  (\ref{presionP}) for  the pressure $P$,  the requirements that   pressure be  regular and positive everywhere within the fluid distribution, and  the S.E.C. condition, it follows   the restriction $\tau>8/3$.

The expressions for $p_{\pm}$, $p_{xx}$ and $p_{zz}$ for our model are
\begin{eqnarray}
p_+&=&\frac{1}{r_{\Sigma}^2}\left\{ -(1-y^2)\frac{\hat g_{yy}}{s^2}+\frac{\hat g_s}{s}\left[ -2+\frac{s^2}{\tau \mu}(5\mu-3 P)\right]-\left( 1-2\frac{s^2}{\tau}\right)\hat g_{ss}\right\},\nonumber \\
p_-&=&\frac{1}{r_{\Sigma}^2}\left\{ -\frac{\partial_y\left[ (1-y^2)\hat g_{yy}\right]}{s^2}+3\frac{\hat g_s}{\mu}\frac{s}{\tau}(P+\mu)-\left( 1-2\frac{s^2}{\tau}\right)\hat g_{ss}\right\},\nonumber \\
p_{xx}&=&\frac{-1}{r_{\Sigma}^2}\left\{y\frac{\hat g_{y}}{s^2}+\frac{\hat g_s}{s}\left[1+\frac{s^2}{\tau \mu}(3P-\mu)\right]\right \},\nonumber \\
p_{zz}&=&\frac{1}{r_{\Sigma}^2}\left[ -(1-y^2)\frac{\hat g_{yy}}{s^2}+y\frac{\hat g_{y}}{s^2}+\frac{\hat g_s}{s}\left( -1+4\frac{s^2}{\tau}\right)-\left( 1-2\frac{s^2}{\tau}\right)\hat g_{ss}\right],\nonumber \\
\label{laspes}
\end{eqnarray}
where $y\equiv \cos\theta$, $s\equiv r/r_{\Sigma}$, and subscripts denote derivatives.

In order to specify further the function $\hat g$ satisfying conditions (\ref{domains}) we shall assume  the metric function $G$ to be separable, i.e. $G(r,\theta)=H(r)J(\theta)$. Then from (\ref{aygsimpleschw})  it follows that $H(r)$ needs to be at least of order $r^n$ with $n\geq 3$,  and $J(y=\pm 1)=0$ from symmetry conditions.  Using the above, is easy to check that  the energy conditions are satisfied  if we chose the following metric function (with $n\geq 3$)
\begin{equation}
\hat g=r_{\Sigma}^{n+2}(s-1)^2 s^n (1-y^2)\bar J(y),
\label{ghoseng}
\end{equation}
for any arbitrary angular function $\bar J(y)$. Two remarks are in order at this point:
\begin{itemize}
\item The order of magnitude in (\ref{domains}) is $O(1/r_{\Sigma}^2)$, and  hence we only need to require the angular function $J$ to be of order $O(1/r_{\Sigma}^{n+2})$  i.e. $\bar J=\frac{1}{r_{\Sigma}^{n+2}} \hat J$ implying
\begin{equation}
\hat g=(s-1)^2 s^n (1-y^2) \hat J.
\label{ghoseng}
\end{equation}
\item The values of the parameters of the models should be chosen such as to assure the bounds (\ref{domains}) allowed for the functions $p_{xx}$ , $p_{zz}$ and $p_{\pm}$  (see below).
\end{itemize}

We can now proceed to find the specific expressions for the components of the energy--momentum tensor corresponding to the line element obtained above. For doing that we shall consider a specific  choice of function  $\hat J$, namely
	\begin{equation}
	\hat J=\epsilon(1-\kappa y^2),
	\label{jota}
	\end{equation}
where  $\epsilon$ is a parameter defining the oblateness of the source (see below  at  subsection $3.3$), and $\kappa$ is a constant. 
After some lengthy but simple calculations we obtain

\begin{eqnarray}
-T_0^0&=&\frac{e^{-2\hat g}}{8\pi\tau r_{\Sigma}^2}\left\{ 6-\epsilon s^{n-2}\left[ (1-y^2)(1-\kappa y^2)R^{(n)}(s)+\right.\right.\nonumber \\
&+&\left. \left.2\tau (s-1)^2(-8\kappa y^4+2y^2(1+4\kappa)-1-\kappa)\right]\right \},\\
T_1^1&=&\frac{e^{-2\hat g}}{8\pi\tau r_{\Sigma}^2}\left\{ 6\left(\frac{\sqrt{\tau-2s^2}-\sqrt{\tau-2}}{\sqrt{3\tau-2}-\sqrt{\tau-2s^2}}\right)+\right. \nonumber \\
&-&\epsilon s^{n-2}(s-1)\left[2\tau y^2(s-1)(1+\kappa-2\kappa y^2)+\right.\nonumber \\
&+&(1-y^2)(1-\kappa y^2)(s(2+n)-n) \times \nonumber \\
&\times&\left. \left.\left(\frac{\sqrt{\tau-2s^2}(\tau-4s^2)-3\sqrt{\tau-2}(\tau-2s^2)}{\sqrt{3\tau-2}-\sqrt{\tau-2s^2}} \right)\right]\right\},  \\
T_2^2&=&\frac{e^{-2\hat g}}{8\pi\tau r_{\Sigma}^2}\left\{ 6\left(\frac{\sqrt{\tau-2s^2}-\sqrt{\tau-2}}{\sqrt{3\tau-2}-\sqrt{\tau-2s^2}}\right)+\right. \nonumber \\
&+&\epsilon s^{n-2}(s-1)\left[2\tau y^2(s-1)(1+\kappa-2\kappa y^2)+\right.\nonumber \\
&+&(1-y^2)(1-\kappa y^2)(s(2+n)-n) \times \nonumber \\
&\times&\left. \left.\left(\frac{\sqrt{\tau-2s^2}(\tau-4s^2)-3\sqrt{\tau-2}(\tau-2s^2)}{\sqrt{3\tau-2}-\sqrt{\tau-2s^2}} \right)\right]\right\},  \\
T_3^3&=&\frac{e^{-2\hat g}}{8\pi\tau r_{\Sigma}^2}\left\{ 6\left( \frac{\sqrt{\tau-2s^2}-\sqrt{\tau-2}}{\sqrt{3\tau-2}-\sqrt{\tau-2s^2}}\right)+\epsilon s^{n-2}\left[ (1-y^2)(1-\kappa y^2)R^{(n)}(s)+\right.\right.\nonumber \\
&+&\left. \left.2\tau (s-1)^2(-8\kappa y^4+2y^2(1+4\kappa)-1-\kappa)\right]\right\}, \\
T_1^2&=&\frac{e^{-2\hat g}}{8\pi\tau r_{\Sigma}^2}s^{n-3} y \epsilon (s-1)\sqrt{1-y^2}\left\{ (s(2+n)-n)+\right. \nonumber \\
&+&\left. 2(s-1)(1+\kappa-2\kappa y^2)\left[\frac{-\sqrt{\tau-2s^2}(\tau-4s^2)+3\sqrt{\tau-2}(\tau-2s^2)}{(\sqrt{3\tau-2}-\sqrt{\tau-2s^2})(\tau-2s^2)} \right]\right \},\nonumber\\
\end{eqnarray}
with the notation
\begin{eqnarray}
R^{(n)}(s)&=&(\tau-4s^2)(s-1)[s(n+2)-n]+(\tau-2s^2)S^{(n)}(s), \\
S^{(n)}(s)&=& s^2(2+n)(n+1)-2n(1+n)s+n(n-1),
\end{eqnarray}
and hence
\begin{eqnarray}
R^{(n)}(s)&=&n^2\tau-2s\tau(n^2+2n+1)+s^2[n^2(\tau-2)+2n(2\tau-1)+4\tau]+\nonumber \\
&+& 4s^3(n^2+3n+2)-2s^4(n^2+5n+6).
\end{eqnarray}

In order to check the fullfilment of the energy and pressure conditions of the solutions one can directly use the energy-momentum components or verify the relations (\ref{domains}) with
\begin{eqnarray}
&p_{xx}&=\frac{s^{n-2}(s-1)\epsilon}{\tau r_{\Sigma}^2}\left\{ 2\tau y^2(1+\kappa-2\kappa y^2)(s-1)+\right.\nonumber \\
&+& \left.\left[\frac{\sqrt{\tau-2s^2}(\tau-4s^ 2)-3\sqrt{\tau-2}(\tau-2s^2)}{3\sqrt{\tau-2}-\sqrt{\tau-2s^2}} \right](1-\kappa y^2)(1-y^2)[s(2+n)-n]\right \},\nonumber \\
\\
&p_{zz}&=\frac{-s^{n-2}\epsilon}{\tau r_{\Sigma}^2}\left\{ (1-y^2)(1-\kappa y^2)R^{(n)}(s)+\right. \nonumber \\
&+&\left.2\tau(s-1)^2[-8\kappa y^4+2y^2(1+4\kappa)-1-\kappa]\right \}.\nonumber\\
\end{eqnarray}

It is a simple matter to check that  both numerators in  the above expressions (as well as the one for $p_{\pm}$) are restricted by the corresponding functions $c_i(\tau)$ or $c_s(\tau)$, in the whole range of $s$ and $y$, for many allowed values of the parameters $\tau$, $n$, and the required  values of $\kappa$ and $\epsilon$. As an example we show in the Figure 1 the well behavior  of those functions, within the required domains (\ref{domains}), for a couple of configurations of the parameters of the model.

\begin{figure}[h]
	$$
	\begin{array}{cc}
	\includegraphics[scale=0.3]{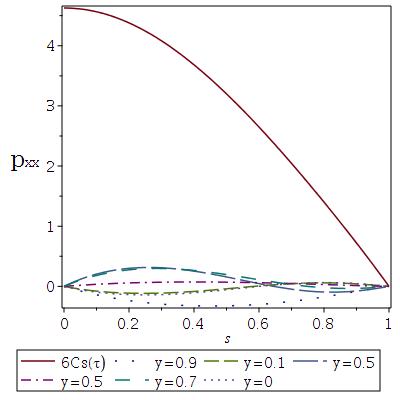}& \includegraphics[scale=0.3]{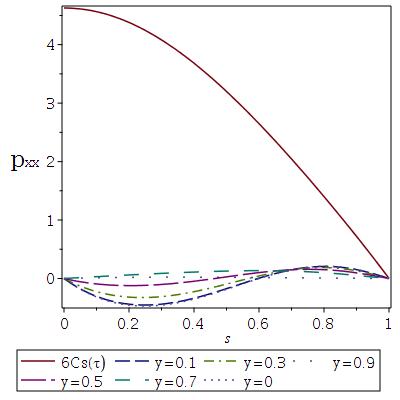} \nonumber\\
	(a) & (b) \nonumber \\ \includegraphics[scale=0.3]{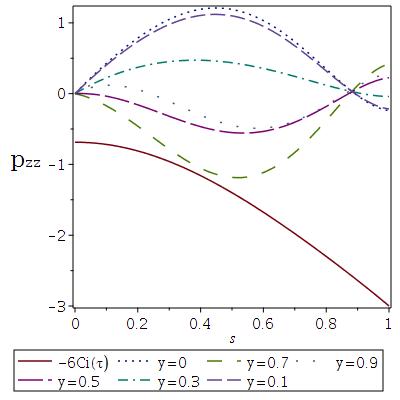}& \includegraphics[scale=0.3]{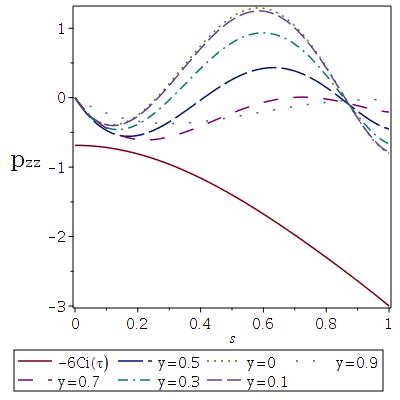} \nonumber\\
	(a) & (b) \nonumber \\
	\includegraphics[scale=0.3]{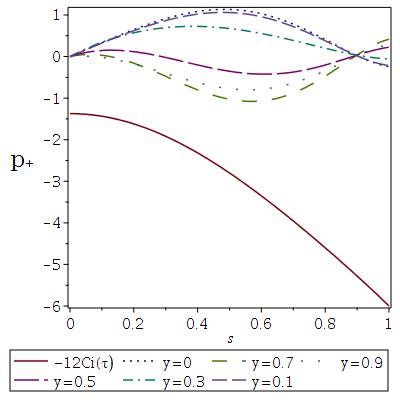}& \includegraphics[scale=0.3]{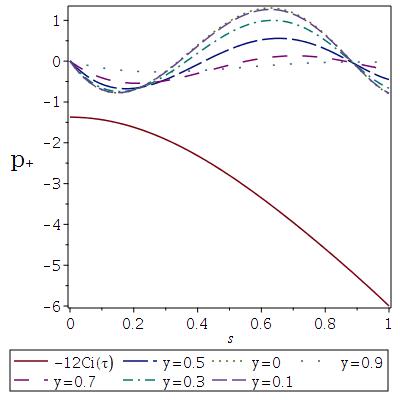} \nonumber\\
	(a) & (b) \nonumber \\ \includegraphics[scale=0.3]{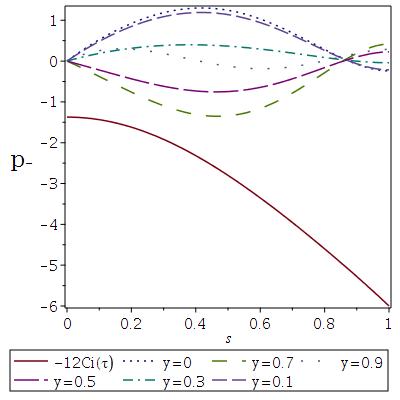}& \includegraphics[scale=0.3]{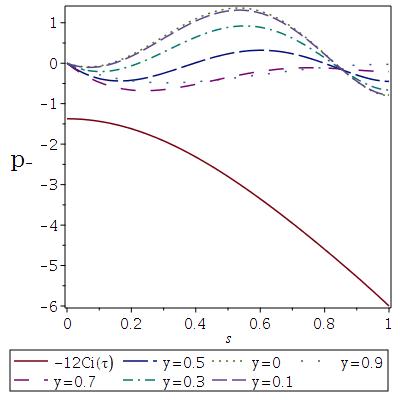} \nonumber\\
	(a) & (b) \nonumber 
	\end{array}
	$$
	{\caption{\label{modelos} {\it Functions $p_{xx}$, $p_{zz}$ and $p_{\pm}$ rescaled by a factor $\tau r_{\Sigma}^2$,  for two different configurations of  the parameters $(\tau ; n, \kappa, \epsilon)$ are represented as functions  of radial variable $s$ and for different values of the angular variable $y$:  (a)  $\tau=2.8$, $n=3$, $\kappa=9$ and  $\epsilon=0.15$ , and  (b)  $\tau=2.8$, $n=3$, $\kappa=1$ and  $\epsilon=0.5$. Solid lines correspond to the upper or lower bounds of the domain for $p_{xx}$, i.e. $6 c_s(\tau)$, $p_{zz}$, i.e. $-6 c_i(\tau)$  and  $p_{\pm}$, i.e. $-12 c_i(\tau)$ all of them  for  $\tau=2.8$. The  non-solid lines in all the graphics  depict functions for different values of the angular variable $y=0, 0.1, 0.3, 0.5, 0.7, 0.9$.} as indicated in the figure.} }
\end{figure}

\subsection{Characterization of the geometry of the source }

With the purpose of providing some information about the ``shape'' of our  source, we shall calculate the proper length  $l_z$ of the object  along the axis  $z$, and the proper equatorial radius $l_{\rho}$, given by
\begin{equation}
l_z\equiv\int_0^{r_{\Sigma}}\frac{e^{\hat g(y=1)-\hat a(y=1)}}{\sqrt A} dz \ , \  l_{\rho}\equiv\int_0^{r_{\Sigma}}\frac{e^{\hat g(y=0)-\hat a(y=0)}}{\sqrt A} d\rho,
\end{equation}
where ${\rho,z}$ are the cylindrical coordinates related to to the  Erez-Rosen coordinates by 
 $\rho= r\sqrt{1-y^2} , \ 
z = r y $.

The expressions above for  the proper lengths   allow us to visualize the flattening of the source with respect to the spherical case, and also to relate it with the  parameter $\tau$ of the source and the free parameter $n$ of the interior metric.

Indeed, in the spherical case ($\hat a=\hat g=0$),  both lengths are identical (as expected)
\begin{equation}
l_z^s=l_{\rho}^s=\int_0^{r_{\Sigma}}\frac{d\xi}{\sqrt{1-p\xi^2}} =r_{\Sigma}\sqrt{\frac{\tau}{2}} \arcsin\sqrt{\frac{2}{\tau}},
\label{lzrho}
\end{equation}
where the fact that  $p=\displaystyle{\frac{2}{\tau r_{\Sigma}^2}}$ has been taken into account and  $l_z^s$, $l_{\rho}^s$ denote the lengths corresponding to the spherical case.

In the general (non--spherical case) we must compare function $\displaystyle{e^{-\hat a(y=1)}}$ with  $\displaystyle{e^{\hat g(y=0)-\hat a(y=0)}}$, since $\hat g(y=\pm 1)$ vanishes along the axis. In addition, our interior metric has been chosen such  that $\hat a=0$ and hence the differences in those lengths are provided by the function
 $\displaystyle{ e^{\hat g(y=0)}}$ which is greater  or smaller than 1 for all values of $s$ in the range $s\in [0,1]$ depending on the sign of the function $\hat J$ on the equatorial plane.

For our line element we obtain for  $l_z$ and $l_\rho$

 \begin{equation}
 l_z=r_{\Sigma} \sqrt{\tau} \int_0^1\frac{ds}{\sqrt{\tau-2s^2}} \quad , \quad l_{\rho}=r_{\Sigma} \sqrt{\tau} \int_0^1\frac{e^{(s-1)^2s^n\hat J(y=0)}}{\sqrt{\tau-2s^2}} ds.
 \end{equation} 
 	 
	 Thus, we see that 	independently on the value   of  $\tau$ and  $n$, this model of source (with $\hat a=0$)  is characterized by an axial length $l_z$ equals to the length of a spherical source $l^s_z$,  and generates a prolate ( $l_z>l_{\rho}$)  source  for negative sign of  $\hat J(y=0)$, whereas an oblate ($l_z<l_{\rho}$)  source follows from a  positive sign of    $\hat J(y=0)$.
	 
	  Since $\hat J(y=0)=\epsilon$, it follows that $\epsilon$ can be related to  the oblateness of the source because $\epsilon>0$ ($\epsilon<0$) implies $l_{\rho}>l_z$ ($l_{\rho}<l_z$) leading to an oblate (prolate) source respectively.
 
Some examples are depicted in Figures 2 and 3.  Thus in  Figure 2 we have chosen  $\hat J=\pm 1$, whereas  in Figure 3  we have $\hat J= 0.8(1-9y^2)$.

\begin{figure}[h]
	$$
	\begin{array}{cc}
	\includegraphics[scale=0.5]{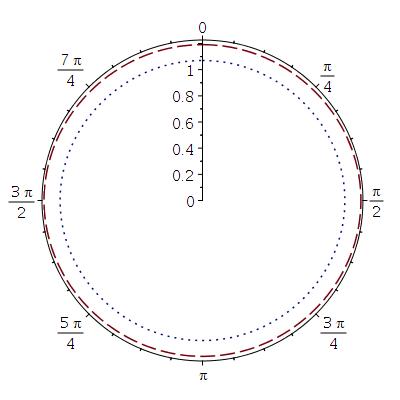}& \includegraphics[scale=0.5]{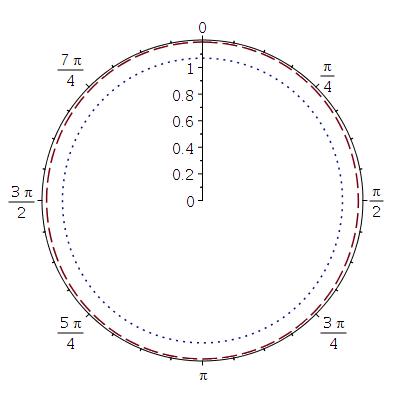}  \nonumber\\
	(a) & (b) \nonumber
	\end{array}
	$$
	\caption{\label{longitudes} {\it$l(\theta)/r_{\Sigma}$  for different values of $\tau$:  (a) with $n=3$, $\kappa=1$ and  $\epsilon=1$ , and  (b) with $n=3$, $\kappa=1$ and  $\epsilon=-1$. Dashed line corresponds to a value of $\tau=2.8$ whereas the pointed line is for $\tau=5.7$. The exterior solid line is a circle.}}
\end{figure}

\begin{figure}[h]
	\includegraphics[scale=0.6]{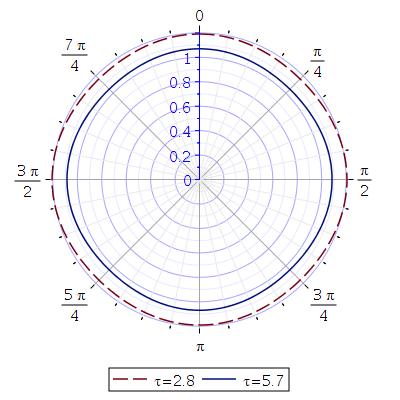}
	\caption{\label{longitudes} {\it  $l(\theta)/r_{\Sigma}$, for different values of $\tau$  with $n=3$, $\epsilon=0.8$ and $\kappa=9$}.}
\end{figure}

It would be convenient to introduce here the concept of ellipticity ($e$), which in terms of  $l_z$ and  $l_{\rho}$, is defined as $e\equiv 1-\frac{l_{\rho}}{l_z}$. The two extreme values of this parameter are $e=0$, which corresponds to a spherical object, and   $e=1$ for the limiting  case when  the source is represented by a disk. In between of these two extremes we have  $e>0$ for a  prolate source and  $e<0$ for an oblate one.

Figure 4 shows the ellipticity $e$  of the source as a function  of the parameter $\tau$, for different values of the  parameter $n$. For positive (negative)  $J$ the ellipticity is  negative (positive) corresponding to an oblate (prolate) source. As can be seen, the relation between $|e|$ (the absolute value) and $\tau$ for any value of $n$ shows that   greater  is $\tau$   greater  is $|e|$. It is also observed from the Figure 4 that the deformation of the source with respect to the spherical case, for any fixed value of $\tau$, is smaller for larger values of  $n$. It also must be pointed out that the dependence of ellipticity on the angular form of the metric function $\hat g$ is weak, since $e$  relates  the proper length at the axis ($y\pm 1$) and the equatorial plane ($y=0$), and so $\hat J(y=0)$ barely provides a major difference in $e$ for different functions $\hat g$.

\begin{figure}[h]
	$$
	\begin{array}{cc}
	\includegraphics[scale=0.34]{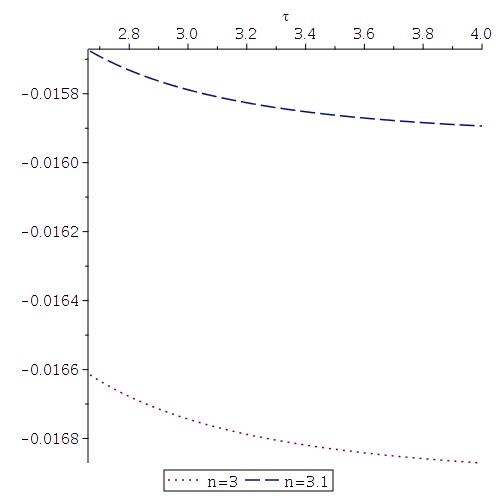}& \includegraphics[scale=0.4]{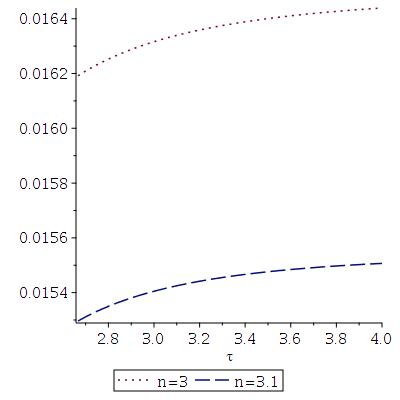}  \nonumber\\
	(a) & (b) \nonumber
	\end{array}
	$$ 
	\caption{\label{ellip} {\it The ellipticity as function of  $\tau$, for different values of  $n$ with $\kappa=1$ and  (a) $\epsilon=1$ and (b) $\epsilon=-1$}}
\end{figure}

\subsection{The complexity of the source}

In recent papers \cite{com1,com2} a new definition of complexity for self--gravitating 	fluids has been proposed, which has been proved to be particularly  suitable for measuring the degree of ``complexity'' of a given fluid distribution. The proposed definition is based on a set of scalar variables called complexity factors, appearing in the orthogonal splitting of the Riemann tensor. More specifically these scalars determine the trace--free electric part of the Riemann tensor.  In the spherically symmetric case \cite{com1} there is only one such a scalar, whereas in the most general axially  symmetric case \cite{com2} there are three of them. 

The electric part of the Riemann tensor $Y_{\mu \nu}=R_{\mu \alpha \nu \beta}V^\alpha V^\beta$, where $V^\alpha=(1/Z,0,0,0)$ denotes the four--velocity of the fluid in the comoving frame and $R_{\mu \alpha \nu \beta}$ is the Riemann tensor, may  be written as

\begin{equation}
	Y_{\mu \nu}=\Psi (g_{\mu \nu}+V_\mu V_\nu)+\frac{Z^{\prime}}{Z} Y^{TF}_{\mu\nu},
	\end{equation}	
where $3\Psi$ is the trace of the tensor $Y_{\mu \nu}$, and the tensor $Y^{TF}_{\mu\nu}$ is the trace free part of $Y_{\mu \nu}$, whose expressions for our solution 

\begin{equation}
\hat g=(s-1)^2 s^n (1-\kappa y^2) \epsilon,
\end{equation}
with $n\geq 3$, are
	\begin{eqnarray}
	\Psi&=&\frac{Z^{\prime}}{Z}\frac Ar e^{-2\hat g},\\
	Y^{TF}_{11}&=&-\hat g^{\prime}, \\
	Y^{TF}_{22}&=&\hat g^{\prime}r^2 A ,\\
	Y^{TF}_{12}&=&\hat g_{,\theta}.
	\end{eqnarray}
	
From the above equations it follows at once

\begin{eqnarray}
	Y^{TF}_{22}&=&-r^2Y^{TF}_{11}A,\\
	Y^{TF}_{12}&=&-\Omega (s)\Lambda(y) Y^{TF}_{11},
	\end{eqnarray}
with ${\displaystyle \Omega(s)\equiv \frac{2r_{\Sigma} s(s-1)}{ns+2s-n}}$ and ${\displaystyle \Lambda(y)\equiv \frac{y}{\sqrt{1-y^2}}\frac{1+\kappa-2\kappa y^2}{1-\kappa y^2}}$, implying that the trace--free part of the electric Riemann tensor  has only one independent component:
\begin{equation}
Y^{TF}_{11}=-\frac{\epsilon s^{n-1}}{r_{\Sigma}}(s-1)(ns+2s-n)(1-y^2)(1-\kappa y^2)
\end{equation} 

Thus, it appears that in spite of the fact that our solution is axially symmetric, the complexity  is  described by a single complexity factor, as in the spherically symmetric case. A similar situation appears in the Szekeres  space--time.

\section{The relativistic multipole moments}

We shall now  obtain some information about our source from the $RMM$ expressed through the variables describing the source. The general theory for doing that has been developed in \cite{whom}, where explicit expressions of $RMM$ in terms of  integrals over the whole space--time has been obtained. These expressions involve integrals denoted by $T_n$ and $S_n^I$  over the space--time filled with the fluid distribution.

Thus, ${\displaystyle T_n=\int_V H_n \rho_T \sqrt{\hat g} d^3\vec x}$  involves the Tolman density $\rho_T$ (describing the material content of the distribution), and ${\displaystyle S_n^I=-\frac{1}{4\pi}\int_V\xi \partial_k\left(\sqrt{\hat g}\hat g^{kj}\partial_j H_n\right) d^3 \vec x}$  where the integration is carried on over  a volume  extended  up to the boundary, and the following notation has been used
\begin{equation}
H_n \equiv \frac{(2n-1)!!}{n!} x^{i_1i_2..i_n} e_{i_1i_2..i_n} \qquad , \qquad \xi \equiv\sqrt{-g_{00}},
\end{equation}
where $e^{i_1i_2..i_n}\equiv (e^{i_1}e^{i_2}...e^{i_n})^{TF}$ with  $e^k$ being the unit vector along the positive direction of the symmetry axis, and $TF$ denoting its  trace free  part, $\hat g^{kj}$ denotes the inverse metric   and $\hat g$ is the determinant of the three-dimensional metric.

As we shall see the only non vanishing $RMM$ is the monopole, in spite of the fact that the source is not spherically symmetric. 

Indeed, for the first three $RMM$ the corresponding expressions  are (see \cite{whom} for details)
\begin{eqnarray}
M_0&=& T_0+S_0^I,\nonumber \\
M_2&=&\frac{-1}{\tau(\tau-2) \beta_2(\tau)}\left[-\frac{M^3}{3}+T_2+S_2^I \right], \nonumber \\
M_4&=&\frac{2M^2\left(1+12\frac{\beta_2(\tau)}{\beta_4(\tau)}\right)}{7 \tau(\tau-2) \beta_2(\tau)}\left[-\frac{M^3}{3}+T_2+S_2^I \right]-\frac{4}{\tau(\tau-2) \beta_4(\tau)}\left[-\frac{M^5}{5}+T_4+S_4^I \right] ,\nonumber \\
\label{formulas}
\end{eqnarray}
where 
\begin{equation}
\beta_n(\tau)\equiv \left[ P_n(x)\partial_x Q_n(x)\right]_{x=\tau-1} \ ,
\end{equation}
$P_n(x)$ and $Q_n(x)$ being the Legendre polynomials and the Legendre polynomials of second kind respectively.

The calculations of the integrals $T_n+S_n^I$ for our model  with the metric  (\ref{interior}) produces, for $n\geq 1$
\begin{equation}
T_n+S^I_n=\frac{M^{n+1}}{n+1}+ \int_0^{r_{\Sigma}} \frac{Z}{2 \sqrt{A}} dr\int_{-1}^1 dy\left[\partial_y(H_n (1-y^2)\hat a_y)\right] 
+ \frac 12 \int_{-1}^1 dy\left[H_n \hat a^{\prime} Z \sqrt{A} r^2\right]_0^{r_{\Sigma}},
\label{churro}
\end{equation}
whereas for $n=0$ we have  $S^I_0=0$ and $T_0=M$.
The second term vanishes when performing the angular integration, and after developing the third term we have

\begin{equation}
T_n+S^I_n=\frac{M^{n+1}}{n+1}+\frac 12 \int_0^1 dy  H_n(r_{\Sigma}) A_{\Sigma} r_{\Sigma}^2 \hat a^{\prime}_{\Sigma}.
\label{churrito}
\end{equation}

We shall  now analyze the expressions above with some detail.  First of all let us notice that if the source is spherically symmetric then
the integral in (\ref{churrito}) can be calculated, producing   $\frac{M^{n+2}}{n+1}\tau(\tau-2) \hat a^{\prime}_{\Sigma}$, and therefore all the multipole moments (\ref{formulas}) higher than monopole  are proportional to $\hat a^{\prime}_{\Sigma}$.	Of course in the non-spherical case  the integral in (\ref{churrito}) should be evaluated in each case since it depends on the angular variable appearing at $\hat a^{\prime}_{\Sigma}$. Next, let us remind that  matching conditions require  $\hat a^{\prime}_{\Sigma}=\hat \psi^{\prime}_{\Sigma}$,
where $\hat \psi^{\prime}_{\Sigma}\equiv (\psi^{\prime}-\psi^{s \prime})_{\Sigma}$  denotes the derivative of the difference (evaluated at the boundary) between the exterior metric function $\psi$ considered and the corresponding one of Schwarzschild $\psi^s$.

Hence, the following  conclusions can be derived from the above comments

\begin{enumerate}
\item Multipole moments higher than monopole vanish if and only if  $\hat a^{\prime}_{\Sigma}$ vanishes, or equivalently iff $\hat \psi^{\prime}_{\Sigma}=0$. This implies that any spherically symmetric  source can only  be matched to the Schwarzschild metric. Although this is a known result, we have proved it  resorting, for the first time as far as we are aware,  to $RMM$.

\item For a general  vacuum non-spherical solution our method provides  a metric function $\hat a$ for the interior metric, whose derivative at the boundary,  in general, is non-vanishing $\hat a^{\prime}_{\Sigma}=\hat \psi^{\prime}_{\Sigma}$, implying that  multipole moments higher than the monopole  are different from zero.

\item Any source,  whether  spherical or not, smoothly matched to the Schwarzschild metric only posses one multipole moment different from zero, which is  just the monopole of the spherical vacuum solution. This conclusion does not depend on how the interior functions $\hat g$ and $\hat a$ are chosen among those allowed.

\end{enumerate}

\section{Conclusions}
We have presented a general approach to obtain non--spherical sources producing  the spherically symmetric Schwarzschild space--time outside the source. A specific solution was completely described, which matches smoothly to  the Schwarzschild space--time on the boundary surface of the fluid distribution, and satisfies all the usual  physical requirements imposed on any physically meaningful source. Besides, some geometric aspects of the source were examined.

The obtained source is  characterized by a single complexity factor,  as it  happens for spherically symmetric fluids, unlike the general axially symmetric case which in general implies three complexity factors. It is worth noticing that another non--spherical source (Szekeres) matchable to the Schwarzschild metric, is also characterized by a single complexity factor. Such a situation tempts us to conjecture that any non--spherically symmetric source matched to Schwarzschild space--time should be characterized by a single complexity factor. The proof of such a conjecture  is of course out of the scope of this work.

We have also analyzed the source in terms of the  $RMM$ expressed through the interior metric. Doing so we have shown that while a non spherically symmetric source may generate a spherically symmetric space--time, the inverse is not true, i.e. any spherically symmetric source  can only be matched to  the Schwarzschild metric.

Finally, let us conclude with some thoughts about the primary motivation of this work. We are well aware of the fact that observational evidence seems to suggest that deviations from spherical symmetry in compact self-gravitating objects (white dwarfs, neutron stars) are likely to be incidental rather than basic features of these systems. This explains why  almost all known models of compact objects serving as sources of the exterior the Schwarzschild metric are assumed to be spherically symmetric as well. However, as we have seen in this work,  there is a wealth of models, more general than spherically symmetric ones, that could be also considered as sources of  Schwarzschild space--time, and which being endowed with a larger number of  degrees of freedom may provide a much wider class of stellar models  encompassing more interesting physical scenarios.

\section*{Acknowledgments}

This  work  was  partially supported by the 
Grant PID2021-122938NB-I00 funded by MCIN/AEI/
10.13039/501100011033 and by “ERDF A way of making Europe”, as well as  the Consejer\'\i a
de Educaci\'on of the Junta de Castilla y Le\'on under the Research Project Grupo
de Excelencia GR234 Ref.:SA096P20 (Fondos Feder y
en l\'\i nea con objetivos RIS3).


\end{document}